# Dynamics of Au (001) Surface in Electrolytes: In-Situ Coherent X-ray Scattering and Scanning Tunneling Microscopy

M. S. Pierce[a†], V. Komanicky[a,b], A. Barbour[a], D. C. Hennessy[a], J.-D. Su[c], A. Sandy[c], C. Zhu[a] and H. You[a]

[a] Argonne National Laboratory, Materials Science Division, Argonne, IL 60439
[b] Safarik University, Faculty of Science, Košice, Slovakia
[c] Argonne National Laboratory, Advanced Photon Source, Argonne, IL 60439

We studied dynamics of Au (001) surface *in situ* in 0.1 M $HClO_4$ electrolyte solution using both coherent x-ray scattering and scanning tunneling microscopy (STM). The surface of Au (001) is known to reconstruct at cathodic potentials; the reconstruction lifts at anodic potentials. In our *in-situ* STM experiments, the measurements focus on time-dependent progressions of surface morphology during slow potential sweep. In our *in-situ* coherent x-ray scattering measurements, we demonstrate that the equilibrium surface dynamics are directly measurable and the measured dynamics are consistent with morphological evolution of the STM images. These experiments represent the first successful application of coherent x-ray scattering to the study of electrochemical interfaces *in situ*.

[†] Currently at Rochester Institute of Technology, Department of Physics Rochester NY 14623

Elucidation of structure and stability of metal surfaces in contact with electrolyte is key to understanding many fundamental and industrials processes such as adsorption of ions and molecules, electrocatalysis, corrosion prevention and electrodeposition/electroplating. Yet complete knowledge of surface structures, in particular their dynamics, is still to be investigated. Here we present a new x-ray tool that can provide dynamic information about the evolution of Au surfaces in electrochemical solution by using surface x-ray photon correlation spectroscopy (SXPCS). This technique is unique in that it provides a quantitative measure of dynamics of the *surface*, even when the sample is held at a constant potential and the average properties of the system remain constant.

Low-index gold surfaces offer a rich variety of surface phases, which are relatively easily prepared and transferred in a clean manner into an electrochemical cell.[1] Gold surfaces also exhibit relatively robust stability in the electrolyte within a limited potential range. Various reconstructed phases and their transformations have been studied by traditional electrochemical techniques such as voltammetry[2] alone or coupled with modern in situ surface probing techniques such as grazing incidence X-ray scattering (GIXS)[3] and scanning tunneling microscopy (STM).[4] In particular, the Au (l00) surface reconstructs into the hexagonal phase, which is stable over a large potential range. When the denser 'hex' phase transforms into square (1×1) phase, as shown in Figure 1, the excess atoms are ejected in the topmost layer and form scattered islands. This transition is fairly rapid under voltammetric conditions for lifting the 'hex' phase. However, the reverse transition from (1×1) to 'hex' phase transition exhibits slow kinetics, requiring ~10 min after jumping well into the hex reconstruction potential in weakly adsorbing electrolytes. It was suggested that it is slow because substantial diffusion of atoms from terrace edges is required to form the denser hex phase in potentiodynamic STM measurements[5] indicating that nucleation and gold adatom mass transport are main factors governing the hex phase growth.

In this paper we show that coherent x-ray scattering can be an alternative and complementary approach to *in-situ*, dynamic STM. Importantly, the time resolution of coherent x-ray scattering will substantially improve as modern x-ray light sources and techniques improve, particularly with continuous operation sources such as an Energy Recovery Linac based facility.[6] Such coherent x-ray techniques can be used not just to investigate phase transitions on Au (100) electrode surfaces under electrochemical conditions, but also to investigate the mobility of phases in equilibrium and access timescales of those processes. In the current experiment, the x-ray data are compared with the images acquired by electrochemical STM (ECSTM) of the Au(001) surface, as well as with earlier electrochemical x-ray scattering experiments to demonstrate the validity of this new technique.

Coherent x-ray scattering differs from the ordinary incoherent x-ray diffraction in its high sensitivity to the structural and temporal details. The coherent fraction of x-rays is typically small and an ensemble sum is necessary to improve counting statistics that tends to wash out the structural and temporal details. In particular, coherent *surface* x-ray scattering (SXS) was doubly difficult because of the inherent intensity loss in SXS to improve the surface sensitivity. In recent years, the coherent fraction, or brilliance, increased dramatically in new synchrotrons. The high brilliance of new synchrotrons improved the chance of successful demonstration of coherent SXS, which was not possible with the older generation synchrotrons. Recently, we developed a coherent SXS technique, the surface-sensitive x-ray photon correlation spectroscopy (SXPCS),[7] that is applicable to electrochemical interfaces.

**Experimental**

To obtain well-prepared, consistent facet surfaces for the experiments, 6 mm gold crystals from Matek GmbH were mounted in a goniometer, oriented to the desired direction by X-ray diffraction, and cast in epoxy resin. The crystals were mechanically polished using sandpaper ranging from 600 to 4000 grit, followed by polishing with 1 μm, 0.3 μm and 0.05 μm alumina suspension. The single crystal surfaces prepared in this way have a miscut typically less than 0.2°. The crystals were then bulk annealed using an RF induction heater setup at ~ 900 ºC for 24 hours in a quartz tube under flow of argon/hydrogen (3% hydrogen, high purity) and typically showed ~ 0.10° to 0.15° mosaic spread at the (002) reflection when checked with an x-ray diffractometer. Subsequent surface annealing was performed in the same heating assembly for 15 minutes at 900 ºC in argon/hydrogen flow. After the crystal had cooled, a droplet of ultrapure water was then placed on the gold surface while still in the argon/hydrogen flow to protect the surface from ambient contamination during transfer.

The crystal was mounted in an x-ray cell or ECSTM cell for subsequent experiments. Platinum counter electrodes were used in both cells. All electrolyte solutions were prepared from J. T. Baker Ultrex II reagents and 18 MΩ cm$^{-1}$ water purified by a Mili-Q reagent water system. STM images were acquired using DI ECSTM system controlled by a Nanoscope IIIa station interfaced with a computer. ECSTM measurements were performed in a home-made Kel-F® cell with a Kalrez® O-ring pressed against the working electrode surface to seal the cell. Pt-Ir STM tips were insulated with nail varnish and left to dry overnight prior to use.[8] A home-made Ag/AgCl reference electrode was used for the ECSTM experiments. For x-ray measurements a low leaking Ag/AgCl (KCl sat.) reference electrode from Bioanalytical Systems Inc was used. Synchrotron SXS experiments were performed in standard reflection cell made from Kel-F® in 0.1 M perchloric acid. All potentials in the text are stated with respect to this reference.

Coherent x-ray surface scattering was performed at the beamline 8ID in reflection geometry. A single bounce Si(111) monochromator was used to set the photon energy to 7.36 keV. Precision slits were placed in front of the sample and narrowed to 10×10 μm$^2$ to achieve the desired transverse coherence. With the slits narrowed, incident flux on the sample was ~10$^9$ photons/sec. We typically detected ~$10^2$ photons/sec after scattering from the surface and through the electrochemical solution at the anti-Bragg condition, the point of the weakest intensity but most surface-sensitive position in the specular rod.

## Results and Discussion

### Scanning Tunneling Microscopy Results

Unlike an Au(001) surface in vacuum, even when we begin with the well annealed surface, the hex reconstruction in electrochemical solution tends to remain in patches and islands after a few potential cycles due to the lack of thermal annealing that can restore the long-range reconstruction. Examples of such topographic (not atomic) images are shown in Figure 1 at 3 different potentials. The islands appear much more coarsened when the reconstruction is lifted at 600 mV, while they appear as smaller patches at 0 mV. However, upon closer examination, we find that the shapes of the islands appear round-edged at 0 mV with hexagonal symmetry, but square-edged at 600 mV. The hexagonally shaped islands do not appear to be aligned with any underlying bulk lattice direction. The change to square topography, where the edges run along the crystal symmetry directions, indicates the internal symmetries of the islands. This coarsening and shrinking behavior repeats with further cycling of the potential. Subsequent voltammetry does not improve the overall morphology, indicating the surface structures have reached a steady state.

The corresponding Fourier transformations (FT) of the three images are also shown to illustrate the symmetries of the topography. For applied potentials of 300 mV and below, the shapes of the islands on the surface have no well-defined edges or orientations. The corresponding FT shows round contours (slightly elliptic probably because of distorted STM images), which is consistent with a large fraction of 60 degree angles with no preferred in-plane orientation. At higher potentials the corresponding FT contours form square or rectangular shapes, indicating that both the islands and the terraces become square-edged and that the edges run generally along the crystal symmetry directions.

Coherent Surface X-ray Scattering Results

Our coherent x-ray scattering measurements contain two kinds of information about changes occurring on the surface. Just as in ordinary SXS, we are capable of measuring potential dependent changes to the overall scattering intensity, albeit much reduced, which provides information regarding the structure and kinetics of the surface. However, the microstate of the surface is recorded in the speckle patterns, and changes in the position and intensity of the speckle is a measure of the real-time dynamics of the surface structure. We will first discuss some of the more general properties of the data collected, then the information provided by the overall intensity, and lastly cover the observed speckle dynamics.

For the potential dependant data, we typically applied a given potential and then collected data for ~ 5000 sec before proceeding to the next potential. Our potential sweeps always proceeded from low to high. Each exposure of the CCD lasted typically 1-5 seconds. An example of the observed scattering is shown in Figure 2 as a function of applied potential. As the potential is increased, there is at first no observed change to the overall scattering, followed by a sharp drop at 300 mV. This observation is also largely consistent with earlier in-situ scattering experiments conducted by Ocko et al.[3]

We observed one additional feature due to our long dwell times at constant applied potentials. We take typically a step of ~ 100 mV and then hold that potential for significant periods of time as we collected data. This was done to ensure the surface was given enough time to equilibrate (and for subsequent data collection once equilibrium was reached). As such the surface spent significantly longer at a given applied potential than is in typical cyclic voltammetry measurements. We were then surprised to find that occasionally the surface reconstruction would begin to lift at potentials as low as +200 mV and +100 mV, much lower than expected. In each of these cases, after significant periods of time (30-45 min) with no changes, the surface began spontaneously lifting the reconstruction within a few minutes. In these cases, the final intensity is recorded as the red squares in Figure 2. The rate of change is consistent with a simple kinetics and an example of the kinetic evolution is given as an inset.

### X-ray Photon Correlation Analysis

The coherent surface x-ray scattering measurements are performed similarly as the incoherent SXS at the specular or off-specular rods. However, instead of the usual point detector such as a scintillator, pixel detectors such as a charge-coupled device (CCD) are used to collect the speckle patterns. The pattern is sensitive to the surface structure and changes as the surface structure changes. The illuminated area is typically limited to ~10 μm across the beam and ~100 μm along the beam by the incidence slits. The limited illumination is necessary to maintain the coherence, and results in much lower overall intensity than ordinary SXS.

Examples of the speckled scattering patterns collected along the specular rod are shown in Figure 3. The scattering at the (001) anti-Bragg condition, while the most sensitive to atomic layer height surface changes, is generally weakest and was not sufficient to perform the experiments directly. However, any changes in response to electrochemical solution or applied potential will be restricted to the surface region of the scattering volume. Because the bulk is effectively static, we can collect speckle dynamics along the specular rod at positions where the total intensity is greater, and yet retain the surface sensitivity.

It is worth noting the presence of significant structure at (0 0 1.95) near the (002) Bragg peak. It means that the mosaic of the Au (001) is small compared to the illuminated area and there is a mosaic distribution within it. However, the (002) speckles do not directly represent the mosaic distribution as is in the ordinary incoherent diffraction. Rather, they are the interference pattern of the mosaic distribution.

The CCD images of the speckles are elongated along the direction of the 2θ scattering angle because the angle of the CCD surface crosses the Ewald's sphere with an angle of θ. Therefore, assuming that the width of the rod in the momentum space is uniform along the rod, the CCD image are longer at smaller angles and the length is proportional to $\sqrt{(2c)^2-(\lambda L)^2}/L$, where c, λ, and L are the lattice constant, wavelength of x-rays, and the reciprocal lattice unit, respectively.

Due to the highly coherent x-rays, detailed information about the microstate of the system is recorded in the speckled diffraction patterns. As the microstate changes, the positions and intensities of the speckles will change in response. This can be seen in Figure 4, in which the scattering from a single 1-d slice of a speckle pattern is plotted as a function of time. The horizontal streaks seen in the right frame indicate approximately the correlation time of the speckle, which can be as long as tens of minutes.

To obtain quantitative measurements of the evolution rates, we calculate an auto-correlation from a given data set. Prior to correlation, backgrounds were subtracted from the images to eliminate dark-current and camera burn artifacts. The normalized auto-correlation was then calculated on a pixel by pixel basis from

$$g_2(\Delta t) = \frac{\langle I(t_0)I(t) \rangle}{\langle I(t_0) \rangle \langle I(t) \rangle},$$

where $\Delta t = t_0 - t$, using a symmetric normalization scheme.[9] Auto-correlations from single pixels were then averaged together with the neighboring pixels. Only pixels with sufficient intensity were used in the average, typically at least 3 photons/pixel, and the dynamic range of the pixel is limited to $10^2$ because of the saturation level of a CCD pixel. A lower-level discrimination was applied for any individual pixel with less than 50 CCD counts (1 photon = 780 CCD counts) being set to 0 to prevent anomalous dark currents from artificially increasing the correlation. Additional details of the correlation analysis can be found elsewhere.[10]

The correlation time is a measure of how long it takes for the illuminated area to reconfigure once on average, locally transforming between the triangular and square symmetries, or growing islands. For low potentials we found typically that a single compressed exponential of the form $g_2(\Delta t) = \beta e^{-(\Delta t/\tau)^\gamma} + 1$ was capable of describing the auto-correlation calculations. For potentials of 300 mV or greater two timescales were evident, and required two additional fit parameters: $g_2(\Delta t) = \beta_1 e^{-(\Delta t/\tau_1)^\gamma} + \beta_2 e^{-(\Delta t/\tau_2)^\gamma} + 1$. An independent compression exponent did not significantly increase the statistical quality of the fits. An example of a two timescale fit is shown in Figure 5.

The correlation times were obtained from the measurements conducted while the surface is in equilibrium and the integrated surface scattering intensity is constant. The correlation time ranges from $10^4$ and high $10^3$ seconds for slow dynamics to middle or low $10^2$ seconds for fast dynamics. It is important to note that the correlation times of reconstructed surfaces at low potential are as high as those measured at ~900 K from the reconstructed surfaces in vacuum.[7] This means that at room temperature, the surface atoms are considerably more mobile in solution than in vacuum. The high mobility of surface atoms in solution has well been recognized for a long time, but our measurements in solution place it quantitatively in comparison to that in vacuum.[10]

Since the correlation time is proportional to the surface diffusion or hopping rate, the correlation at a given temperature is given by $\frac{1}{\tau} \propto \nu_1 e^{-E_b/k_B T}$, where $E_b$ is an energy barrier for an atom to detach from a step edge and $v$ is an attempt frequency. Using this simple approximation and an assumption that the attempt frequency of Au atoms depends only on $T$, we can estimate the reduction of the energy barrier from vacuum to electrolyte environments. In Figure 6, we show the correlation time vs. potential. At low potentials (below 0 mV), the correlation time of the hex-reconstructed phase is $2 \times 10^4$ sec, which is smaller by a factor of 100 than that in vacuum.[7] The reduction in $E_b$ ($\Delta E_b$) then can be estimated to be $\ln(10^2)k_B T$ or ~ 0.14 eV. At the Au (001) and water interface, the water-water bonds at the interface are broken and replaced by Au-H$_2$O bonds. This means Au atoms, with which nothing interacts in vacuum, will be constantly interacting weakly with water molecules and $\Delta E_b$ is the Au-water bond energy. Indeed, it is in reasonable agreement with the Au-H$_2$O bond energy (0.15 eV) estimated by a density functional theory (DFT) study.[11] It is probably reasonable to assume that it will be difficult for two water molecules to bond simultaneously to one Au atom, even though there are twice as many water molecules as Au atoms at the interface.

The correlation times also change dramatically in response to the applied potential, as shown in Figure 6. Focusing only on the squares, we see that the correlation time decreases by a factor 50 over 300 mV change in potential. This compares well with a dramatic increase of the island decay rates in sulfuric acid.[12] Even though we do not expect that perchlorate adsorbs specifically, the change of water dipole might be sufficient to accelerate the dynamics at high potentials. More importantly, though, the change coincides well with the hex-to-(1×1) transition as we have shown with STM measurements (Figure 1).

It is significant that we observe dynamics not just during the transition, but also at all higher potentials once the reconstruction has lifted. The surface remains in a state of dynamic equilibrium even though the surface phase transition from hex to (1×1) has already occurred. Since the logarithm of the inverse correlation time is proportional to the surface activation energy, in this case the energy of desorption from the step or island edges, the correlation time is fit to a change in activation energy shown as the solid lines. If we assume that the attempt frequency is not very sensitive to the surface symmetry since it is driven by thermal vibration, the factor of 50 corresponds to an additional decrease in the activation energy of ~0.1 eV. Therefore, compared to the vacuum hex phase, the atoms in the (1×1) phase in electrolyte desorb/adsorb 5000 times more frequently at room temperature.

The change in dynamics caused by the potential is confirmed in the STM images shown in Figure 7. Two pairs of images are selected, one at 200 mV and one at 500 mV. The first image of each pair was taken shortly after the potential was changed. The second image of each pair was taken ~10 min later. We can see that the two images taken at 200 mV are largely the same, indicating very slow evolution of the step/island morphology. The two images taken at 500 mV are quite different, which indicates significant dynamics during the same time period. From these observations, it is evident that the structure is changing more quickly at 500 mV, but quantifying how much it is changing becomes more difficult due to STM drift, tip-surface interaction, and the decrease in sampled area. Nonetheless, we can barely recognize the step/island morphology of (b) from (d). It is likely that after more time, the surface at 500 mV becomes unrecognizable from the initial state. The approximate time scale that can be determined from STM measurement agrees well with the correlation time of ~400 sec measured from the x-ray speckles.

There exists another correlation time at high potentials as discussed with Figure 5. This correlation time, shown as open circles in Figure 6, is even faster, approaching the experimental time-resolution limit. The origin of this dynamics cannot be determined at this time and we can only speculate. Since it is much faster than the step motion, we believe it is from the local fluctuations of microstates, in length scales of ~10 to ~100 nm, within the terraces and islands, not from their edges. The microstate fluctuations are a result of fluctuations of local symmetry. Examples of local symmetries are shown as the insets in Figure 6. The cubic symmetry can easily breathe to hexagonal symmetry and vice versa. Note that the average maintains the cubic symmetry even though the symmetries may locally fluctuate. These fluctuations are most likely driven by water and ions in double layers since the system is at room temperature. On the other hand, the hexagonal symmetry at low potentials does not have room for fluctuation due to the close packed structure, which explains why we do not see the fast correlation times at the low potential range.

Ultimately further measurement and analysis is necessary before a complete understanding of the observed dynamics can be obtained. XPCS analysis of data obtained at significantly different q-values could perhaps be used to deduce something akin to a diffusion rate, as done in other studies.

In summary, we have demonstrated that SXPCS measurements yield unique equilibrium dynamics information of Au (001) surface in perchlorate. We believe that this is a novel, powerful technique for electrochemical interface studies. We anticipate that future experiments can be performed with different facets of gold, electrochemical potential range, electrochemical solution with strongly or weakly adsorbing anions. Another application which could prove favorable would be to extend in-situ studies of electro-deposition. Step-flow sublimation studies of Pt at high temperature have already been successful [13] and it seems likely that extension of the technique to growth at the solid-liquid interface. [14] We believe that the technique will become applicable to all area of surface electrochemistry as the brilliance of modern x-ray sources continues to increase.

## Acknowledgments

This work and use of the Advanced Photon Source were supported by the U.S. Department of Energy, Office of Basic Energy Sciences, under Contract No. DE-AC02-06CH11357. The work at Safarik University was supported by Slovak grant VEGA 1/0138/10 and VVCE-0058-007.

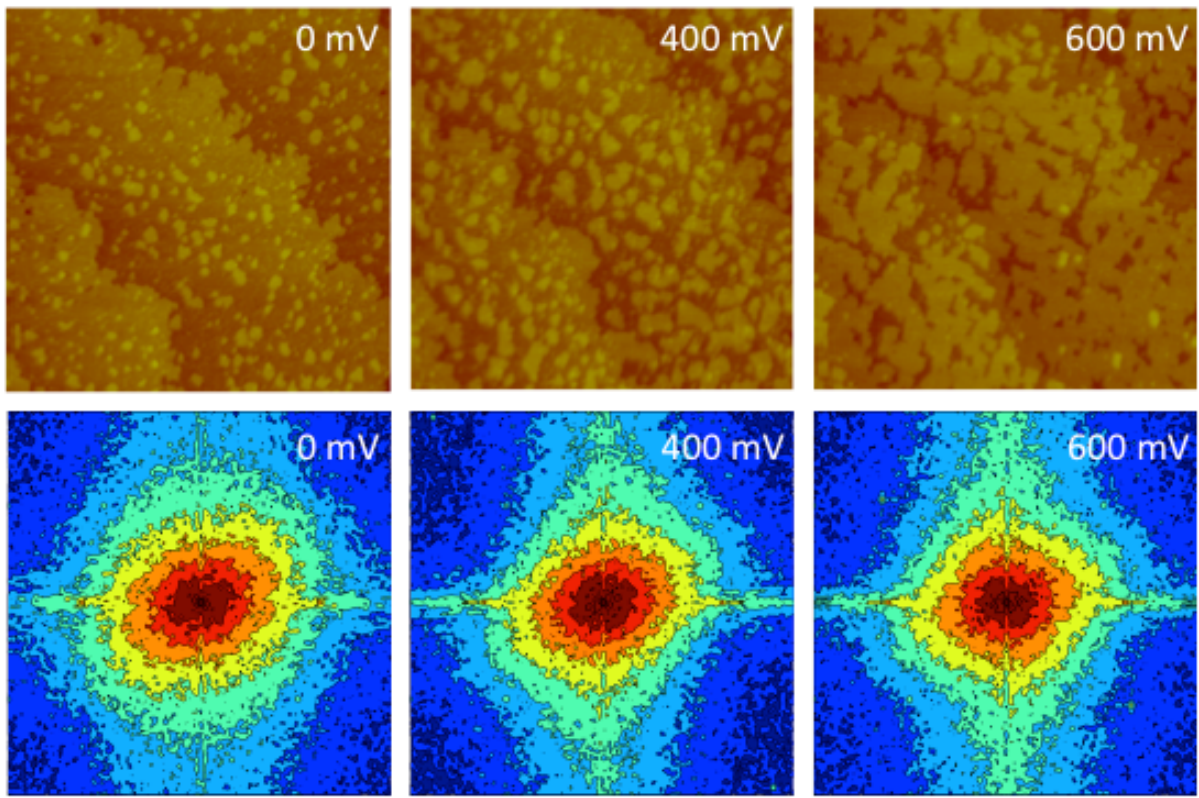

Figure 1. *In-situ* STM images at 3 different potentials, along with the corresponding contour plots of the Fourier transform magnitudes from each image. The STM images are 400×400 nm$^2$. The symmetry of the structure evolves into a square edged state as the potential is increased.

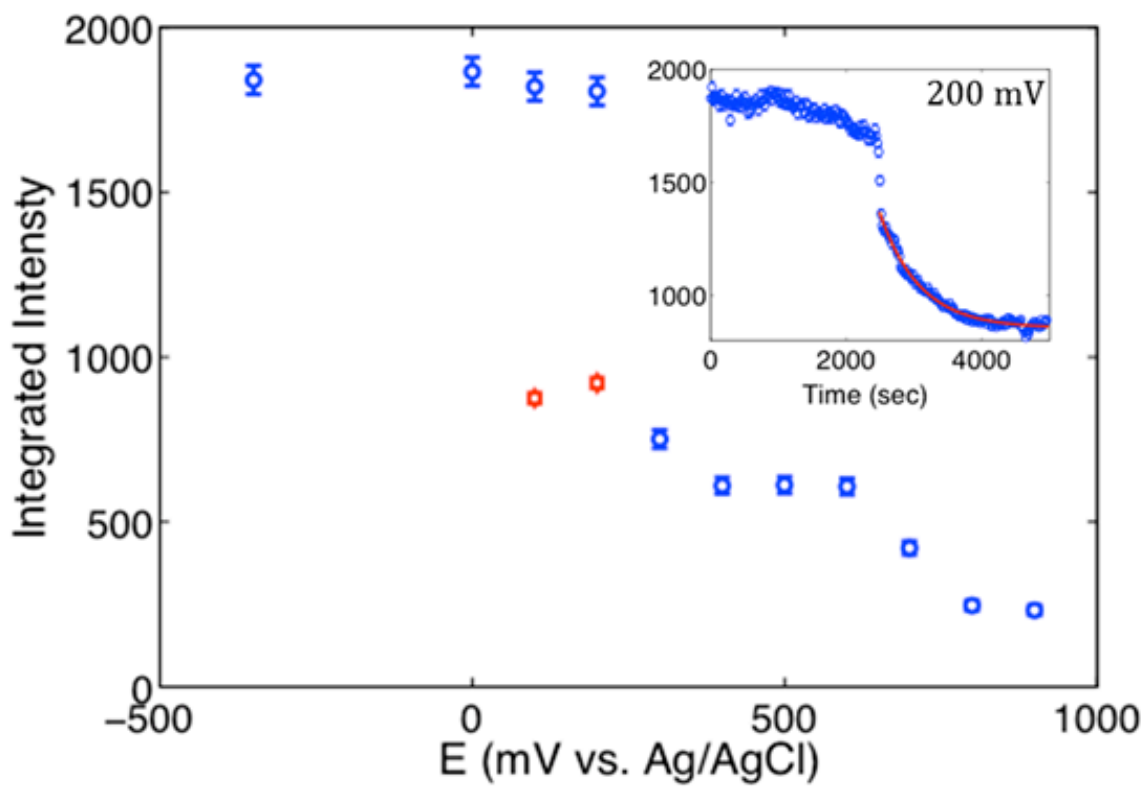

Figure 2. An example of the scattering intensity recorded as a function of applied potential at L = 0.17. The potential sweep started at -350 mV and then increased to +900 mV. The two red squares show the intensity after the spontaneous lifting of hex from metastable equilibrium. Inset shows an example of the metastable equilibrium achieved at 200 mV and the subsequent spontaneous lifting of the hex as a drop in the observed intensity.

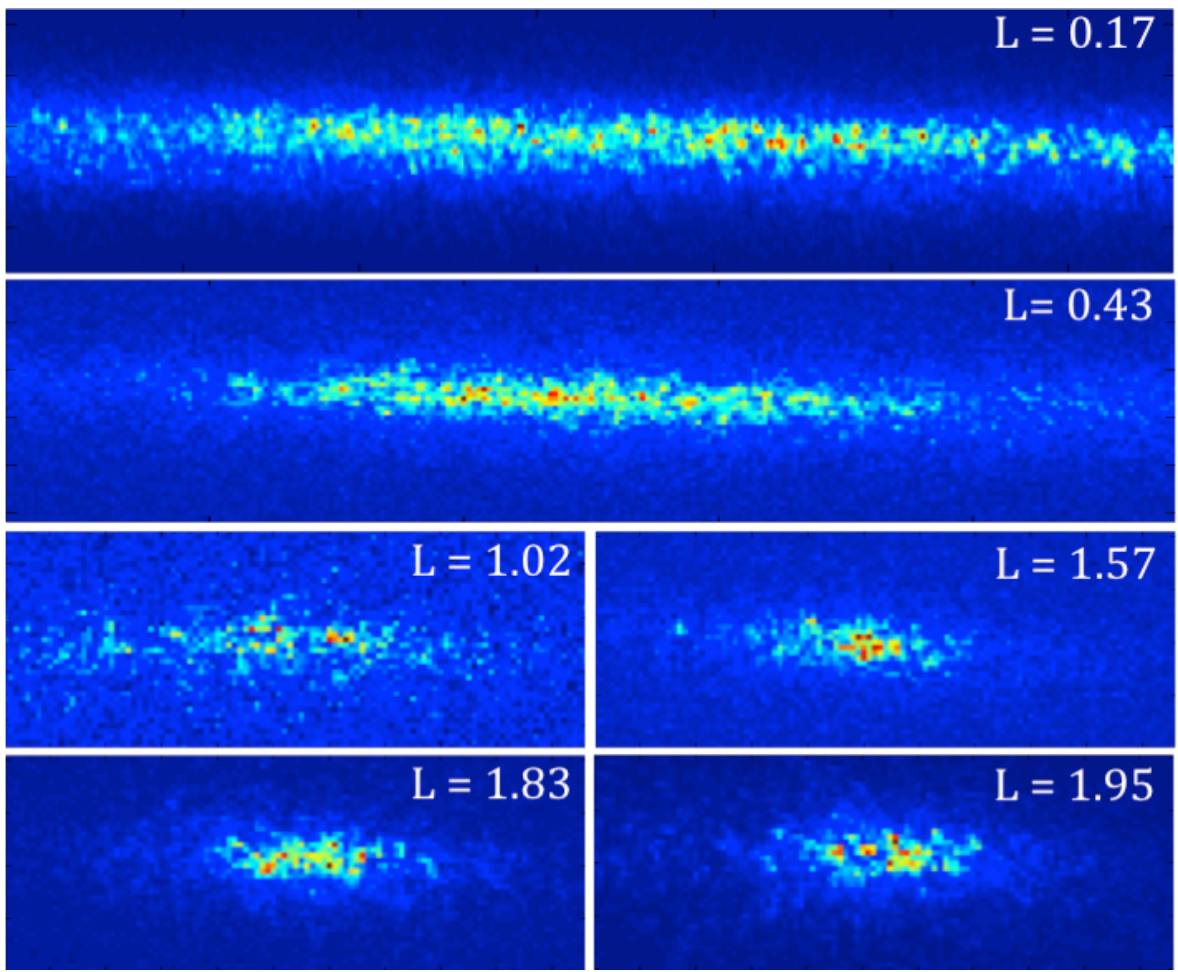

Figure 3. Example speckle patterns collected as a function of L at an applied potential of +100 mV. For small L the image is stretched considerably due to the shallow scattering angles. The scattering angle increases along the horizontal axis in each figure.

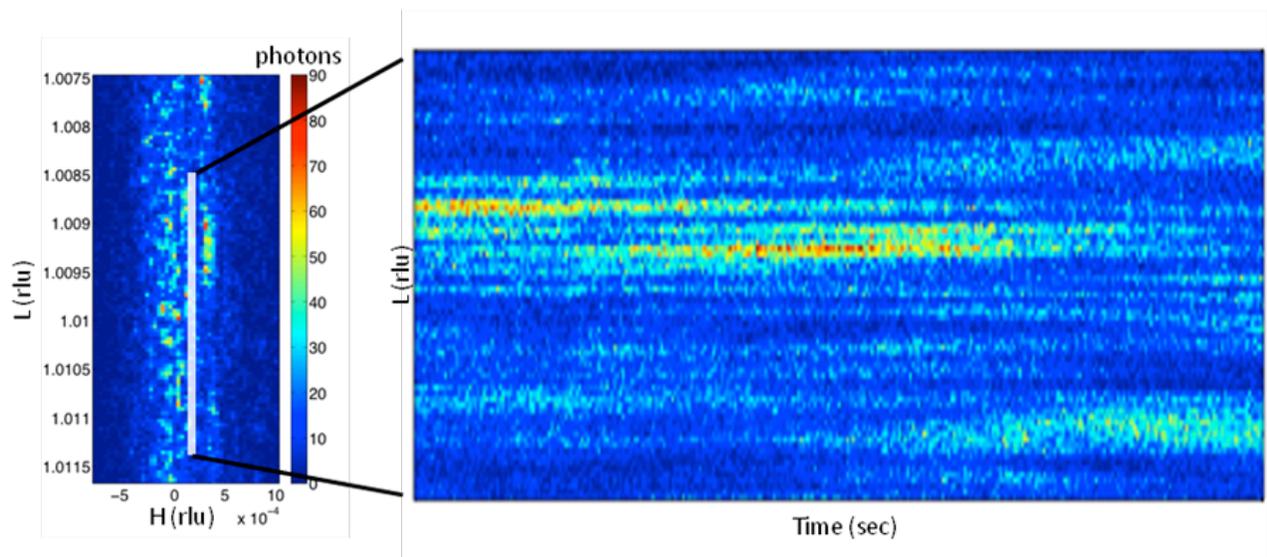

Figure 4. A portion of a single speckle pattern is shown at left. The highlighted 1d region is then plotted vs. time on the right over the course of about 20 minutes. The intensity of the speckle pattern can be seen to change slowly over time as the surface evolves, even though the total intensity remains constant.

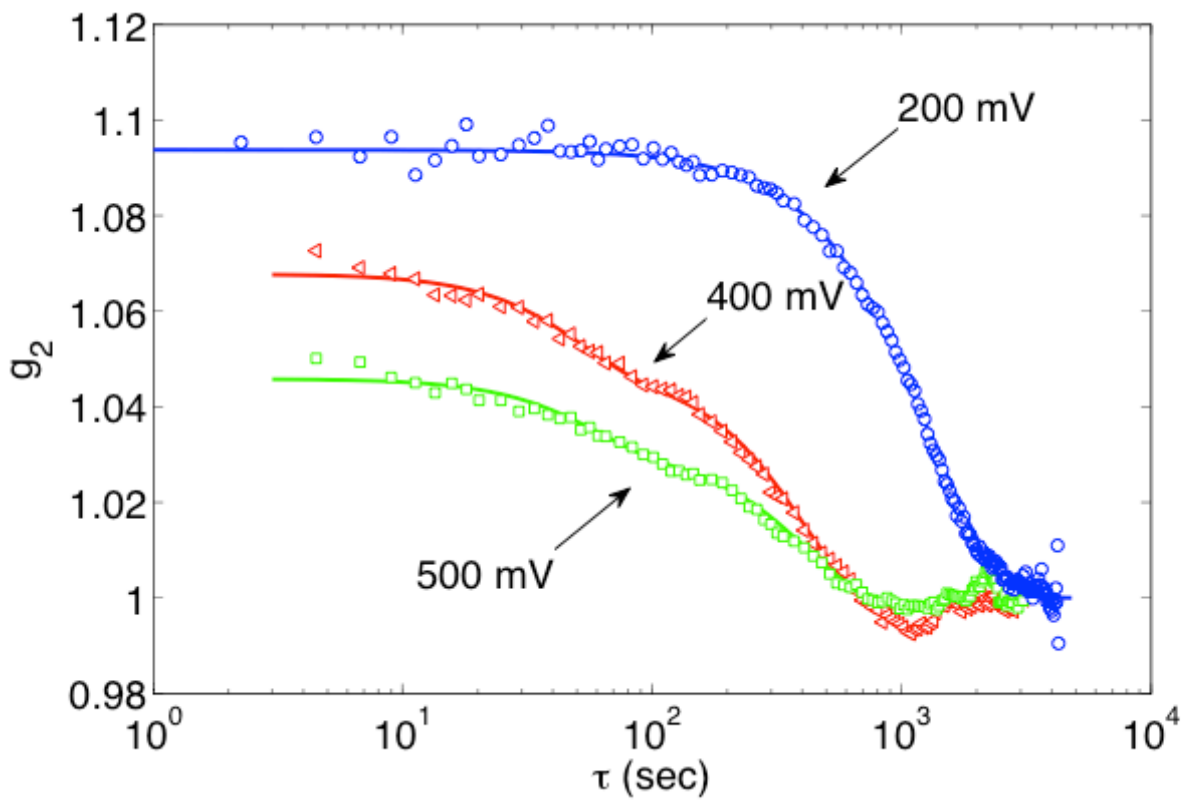

Figure 5. Auto-correlation from the Au (001) surface at 3 different potentials collected at L = 0.17. For potentials at +200 mV and below the decay is governed by a single exponential decay. For higher applied potentials the autocorrelation decays much faster and we see two distinct timescales develop in the autocorrelation functions. Fitting a double exponential returns both time constants. The contrast scale for 500 mV has been offset for clarity.

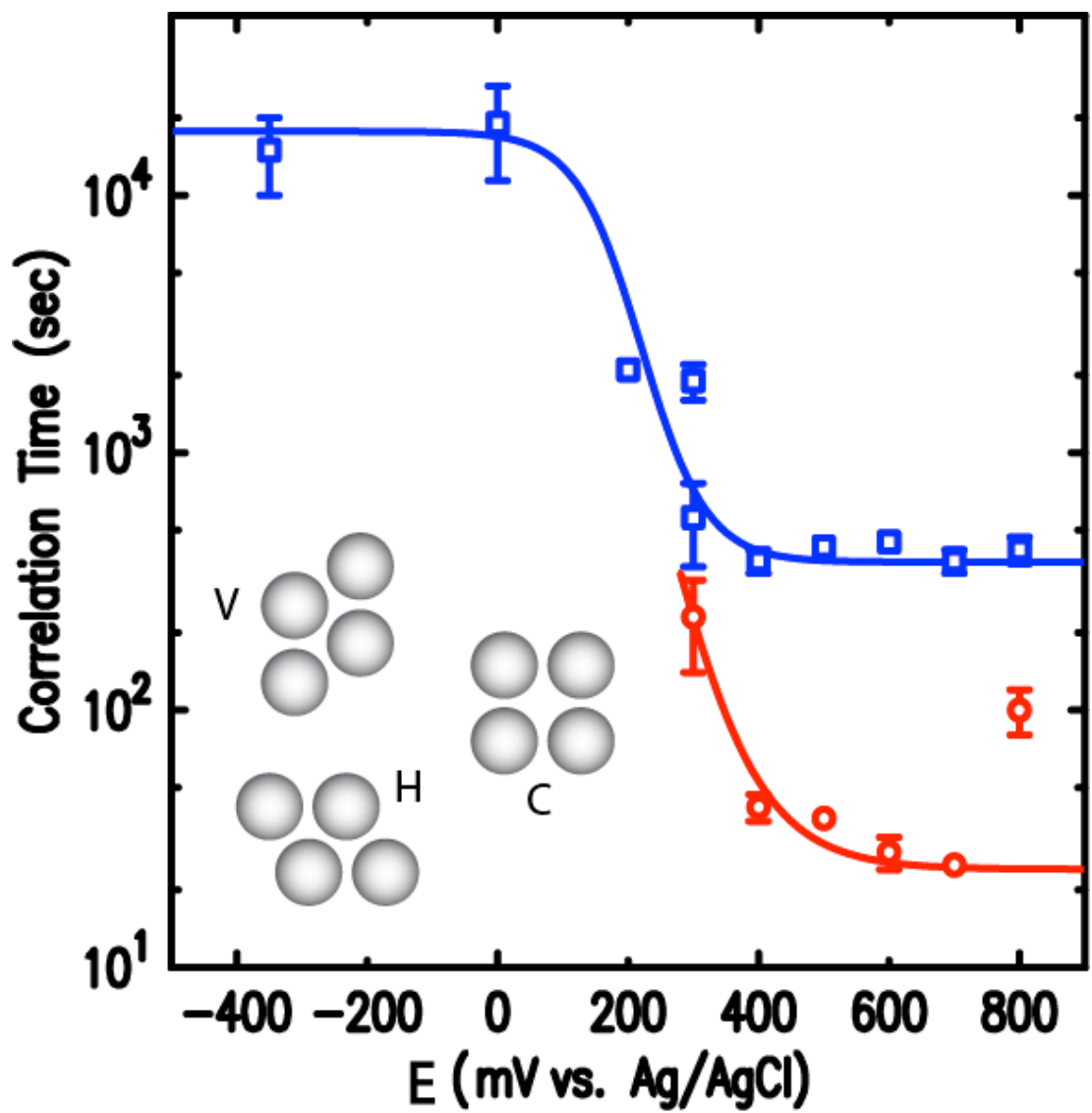

Figure 6. Time constants for Au (001) plotted as a function of applied potential as measured at L = 0.17. For +300 mV and higher, there are two distinct correlation-time constants that are required to fit the data. The solid lines are guides to eyes. The insets are three atomic configurations (see the text for discussion).

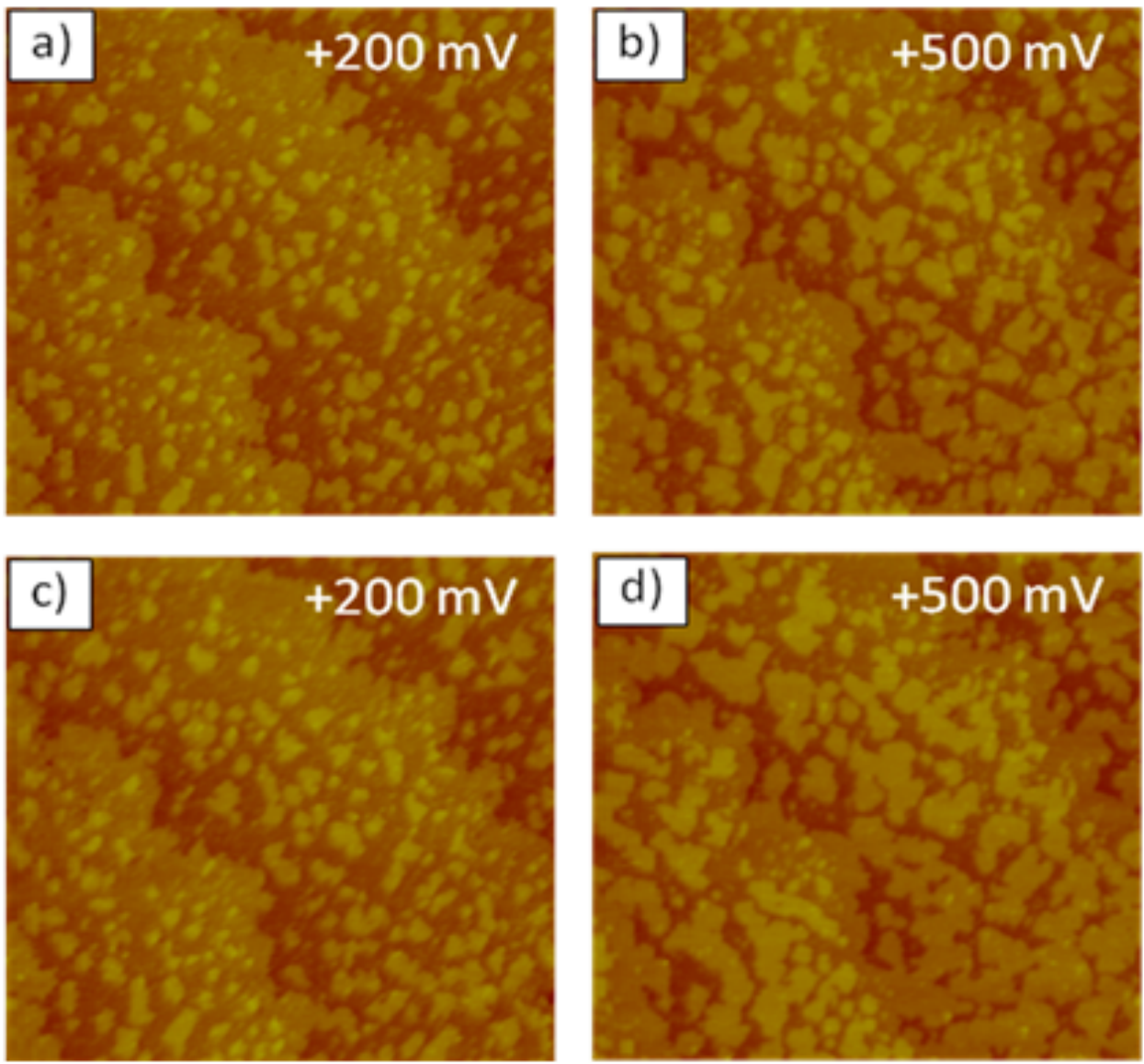

Figure 7. STM images for 200 mV and 500 mV. Each image shows a region of 400×400 nm$^2$. Panels a) and b) show the surface just after the potential had been changed. Panels c) and d) show the same regions 10-15 minutes after being held at a constant potential. It is evident that c) resembles a) more than d) resembles b), but quantifying how much change has occurred is difficult from such images. XPCS offers a quantitative measure of such changes, though in our limited application to the specular rod XPCS alone cannot reveal what or how the surface structure is changing.